\pdfoutput=1
\newcommand{\U}[1]{\,{\rm{#1}}}
\newcommand{\Wcm}{\U{W/cm^2}}
\newcommand{\mf}{\mathbf}
\newcommand{\In}[1]{_{\mathrm{#1}}}

\documentclass[graybox]{svmult}

\usepackage{mathptmx}       % selects Times Roman as basic font
\usepackage{helvet}         % selects Helvetica as sans-serif font
\usepackage{courier}        % selects Courier as typewriter font
\usepackage{type1cm}        % activate if the above 3 fonts are
\usepackage{makeidx}         % allows index generation
\usepackage{graphicx}        % standard LaTeX graphics tool
\usepackage{multicol}        % used for the two-column index
\usepackage[bottom]{footmisc}% places footnotes at page bottom

\makeindex             % used for the subject index
\begin{document}

\title*{Electron correlation and interference effects in strong-field processes}
%\author{M. C. Kohler, C. M\"uller, C.~Buth, A.~B.~Voitkiv,  K.~Z.~Hatsagortsyan,  J.~Ullrich, T.~Pfeifer, and C.~H.~Keitel}
\author{Markus C. Kohler, Carsten M\"uller, Christian~Buth, Alexander~B.~Voitkiv,  Karen~Z.~Hatsagortsyan,  Joachim~Ullrich, Thomas~Pfeifer, and Christoph~H.~Keitel}
\authorrunning{M. C. Kohler \textit{et al.}}
\institute{Markus C. Kohler \at  Max-Planck-Institut f\"ur Kernphysik, Saupfercheckweg 1, 69117 Heidelberg, Germany,\\ \email{markus.kohler@mpi-hd.mpg.de}
\and Carsten M\"uller \at Max-Planck-Institut f\"ur Kernphysik, Saupfercheckweg 1, 69117 Heidelberg, Germany,\\ \email{carsten.mueller@mpi-hd.mpg.de}
\and Christian Buth \at Max-Planck-Institut f\"ur Kernphysik, Saupfercheckweg 1, 69117 Heidelberg, Germany,\\
Argonne National Laboratory, Argonne, Illinois 60439, USA,\\ \email{christian.buth@web.de}
\and Alexander~B.~Voitkiv \at Max-Planck-Institut f\"ur Kernphysik, Saupfercheckweg 1, 69117 Heidelberg, Germany,\\ \email{alexander.voitkiv@mpi-hd.mpg.de}
\and Karen~Z.~Hatsagortsyan \at Max-Planck-Institut f\"ur Kernphysik, Saupfercheckweg 1, 69117 Heidelberg, Germany,\\ \email{k.hatsagortsyan@mpi-hd.mpg.de}
\and Joachim Ullrich \at Max-Planck-Institut f\"ur Kernphysik, Saupfercheckweg 1, 69117 Heidelberg, Germany,\\
Max-Planck Advanced Study Group at CFEL, 22607 Hamburg, Germany,\\ \email{joachim.ullrich@mpi-hd.mpg.de}
\and Thomas Pfeifer \at Max-Planck-Institut f\"ur Kernphysik, Saupfercheckweg 1, 69117 Heidelberg, Germany,\\ \email{tpfeifer@mpi-hd.mpg.de}
\and Christoph H. Keitel \at Max-Planck-Institut f\"ur Kernphysik, Saupfercheckweg 1, 69117 Heidelberg, Germany,\\ \email{keitel@mpi-hd.mpg.de}}
\maketitle

%an abstract (10--15 lines long)
\abstract{Several correlation and interference effects in strong-field physics are investigated. We show that the interference of continuum wave packets can be the dominant mechanism of high-harmonic generation (HHG) in the over-the-barrier regime. Next, we combine HHG with resonant x-ray excitation to force the recolliding continuum electron  to recombine with a core hole rather than the valence hole from that it was previously tunnel ionized. The scheme opens up perspectives for nonlinear xuv physics, attosecond x-ray pulses, and spectroscopy of core orbitals. Then, a method is proposed to generate attochirp-free harmonic pulses by engineering the appropriate electron wave packet. Finally, resonant photoionization mechanisms involving two atoms are discussed which can dominate over the direct single-atom ionization channel at interatomic distances in the nanometer range.}

\section{High-harmonic generation by continuum wave packets}
\label{sec:ccHHG}

High-order harmonic generation (HHG) is a key process in ultrafast science and well-understood within the three-step model~\cite{CORKUM1993}: the bound wave function of an atom is partially freed by a strong laser field, accelerated in it, and driven back to its parent ion.  At that point, the ionized and bound portions of the electronic wave packet interfere, giving rise to a strong, coherent high-frequency dipole response that can lead to the emission of a HHG photon along with the recombination of the electron into the bound state.

In our work \cite{KOHLER2010}, we advance the interference model of HHG~\cite{PUKHOV2003,ITATANI2004} to provide a comprehensive physical picture including continuum--continuum (CC) transitions: any two wave packets of the same electronic wave function that have been split and have acquired different energies  lead to coherent HHG emission when they simultaneously reencounter the core region. The emitted photon energy is exactly the energy difference of the two wave packets.  Starting from this point of view, we found that a new CC transition plays a significant role in the over-the-barrier (OBI) ionization regime. This transition occurs when two wave packets ionized in different half cycles of a laser pulse recollide at the same time.  Note that this transition is different to the CC harmonics described in~\cite{CCHarmonics}.
% lead to a significant HHG response. %The emitted photon energy is exactly the kinetic-energy difference of the two wave packets. 
%Our approach thus provides a fundamental physical of CB HHG to include CC transitions:  wave function splitting or spreading, subsequent simultaneous recollision with different energies, and core-mediated transition results in coherent photoemission at the difference energy.  

\begin{figure}[h]
\centering
\includegraphics[width=7.0cm]{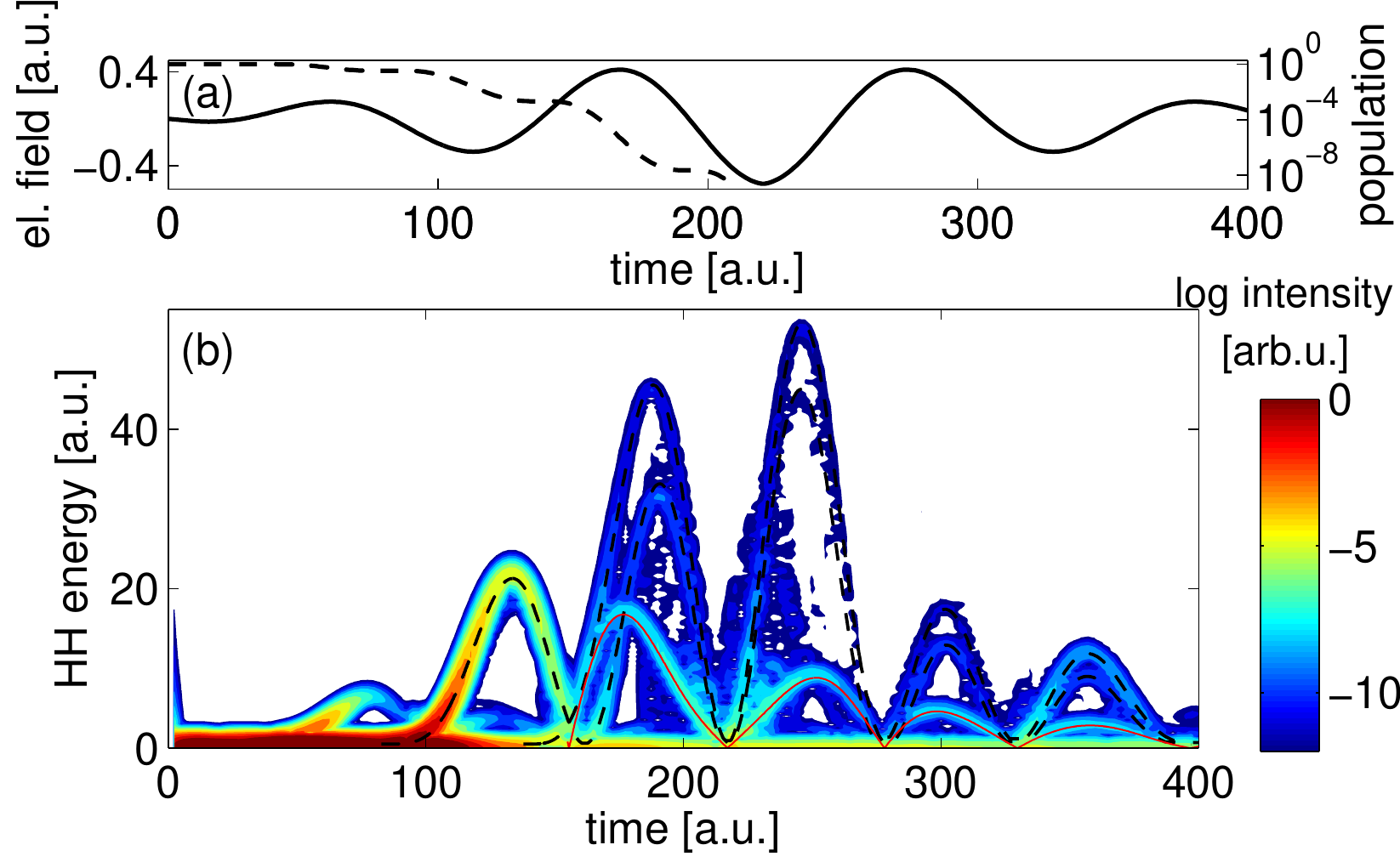}
\caption{\label{fig-HHGnumerical}
Time-frequency analysis of HHG showing the signature of CC wave-packet interference.  a) Laser pulse used for the calculation (solid line, left axis) and ground-state population (dashed line, right axis).  b)~Windowed Fourier transform of the HHG emission.  The dashed black lines are the classically calculated kinetic energies of electrons returning to the ion and the solid red line their difference energy. Reproduced from Ref.~\cite{KOHLER2010}. Copyright (2010) by the American Physical Society.}
\end{figure}

As a realistic model, we study the hydrogen atom in a laser field in the OBI regime and solve the three-dimensional time-dependent Schr\"odinger equation numerically~\cite{BAUER2006}. The laser pulse is shown in Fig.~\ref{fig-HHGnumerical}a and chosen such that almost complete depletion of the ground state occurs on the leading edge of the pulse. 
%This way, CC transitions become the dominating HHG mechanism for large parts of the pulse as shown below. 

To analyze the time-resolved  frequency response of HHG, we calculate the windowed Fourier transform of the dipole acceleration obtained from the TDSE calculation and display it in Fig.~\ref{fig-HHGnumerical}b. For comparison, the two  dashed black lines in the figure are the classical recollision energies for trajectories starting from two different laser half cycles, respectively, which are in agreement with the traditional continuum--bound (CB) signal whereas the red line denotes their difference.
The CC transition is evidenced by the excellent agreement of the quantum-mechanical response with the red line. Interestingly, the CC component of the dipole response is the dominant contribution for several half cycles after $t=150$~a.u. (atomic units are employed throughout). This can be understood from the fact that depletion of the ground state occurs around that time. Then, coherent HHG can only occur by the presence of the various parts of the wave function in the continuum.

%The significance of the signal is pointed out by estimating the photon yield in an energy window between 12~a.u. and 18~a.u. Assuming a typical phase-matched gas volume (radius $100~\mu$m, length $1$~mm and density of $10^{18}$/cm$^3$), we calculate a yield of $10^3$ photons per shot from CB HHG and a photon number of $10^{-1}$  from CC HHG. The typical kHz repetition rates are expected to allow for a measurement of this effect.

Moreover, we developed a strong-field approximation  model for CC HHG suitable for the OBI regime and based on the evaluation of the dipole acceleration  
$\mf{a}(t)=-\langle \Psi ,t\vert \mf{\nabla} V\vert \Psi ,t\rangle$~\cite{Gordon2005} rather than the dipole moment to include the distortion of the recolliding waves by the Coulomb potential required for momentum conservation. The saddle-point approximation is applied to the expression making a computationally fast evaluation of the process possible.

This analytical model also allows to extract the HHG emission phase  $\phi=S(\mf{p}'',t,t'')-S(\mf{p}',t,t')+I_p(t'-t'')-\omega_{\mathrm{H}} t$ with the classical action $S$~\cite{lewenstein}, canonical momentum~$\mf{p}$, HHG emission time $t$, ionization times $t'$ and $t''$, HHG frequency $\omega\In{H}$ and ionization energy $I\In{p}$.
The expression can be converted to  $\phi(I)=\alpha\In{cc}I$ being a function of the laser pulse peak intensity $I$, which reveals a striking difference for CB as compared to CC transitions: the sign of $\alpha\I{CC}$ and $\alpha\In{CB}$ differs for both types of transitions allowing to separate the CC from the CB harmonics via phase-matching. 
This way, the measurement of CC spectra could by employed for qualitatively advancing tomographic molecular imaging~\cite{ITATANI2004}: instead of probing the orbital shape of the active electron, the effective atomic or molecular potential mediating the transition between the two wave packets could be assessed.

%The mechanism of CC HHG as described here can also be understood in close analogy to the free-electron laser (FEL) process, but on an atomic and sub-optical-cycle scale.  This knowledge might enable ultra-compact FEL-like sources in the future.

\section{Emergence of a high-energy plateau in HHG from resonantly excited atoms}
\label{sec:rabiHHG}
In the last two decades, HHG in the non-relativistic regime  has been developed to a reliable source of coherent extreme ultraviolet (XUV) radiation. Its  advancement into the hard x-ray domain would allow for a much wider range of applications. The straightforward approach to employ larger laser intensities is demanding due to the relativistic electron drift~\cite{drift} and the large electron background that is generated by the strong laser field causing phase-mismatch~\cite{Popmintcheva:PM-09}. On the other hand, the large scale x-ray free electron lasers (XFEL) routinely generate several keVs of photon energy but are limited in coherence and, thus, sub-femtosecond pulses with this technique are not in reach at the moment. 
We show~\cite{buth:ra-11} that by combining both, the HHG process and radiation from an XFEL, coherent light pulses can be obtained having the extremely short time structure of the HHG  and photon energies larger than the XFEL. 
In addition to the increase of the HHG photon energy, the scheme can be employed for ultrafast time-dependent imaging~\cite{morishita2008} involving inner shells and for the characterization of the x-ray pulse of the XFEL.

% \begin{figure*}[ht]
% \begin{center}
% \includegraphics[width=0.7\textwidth]{images/rabi_HHG_schematic}
% \caption{Schematic of the HHG scenario as a three-step process: (a) the valence electron is tunnel ionized; (b) the additional high-frequency light excites the core electron; (c) the continuum electron recombines with the core hole.
% } \label{fig:rabi-schematic}
% \end{center}
% \end{figure*}

%A schematic of the proposed scheme is shown in Fig.~\ref{fig:rabi-schematic}.  
The proposed scheme works as follows: atoms are irradiated by both an intense optical laser field and an x-ray field from a FEL.  The x-ray energy is chosen to be resonant with the transition between the valence and a core level in the cation. As soon as the valence electron is tunnel ionized by the optical laser field, the core electron can be excited to the valence vacancy. Then the continuum electron, returning after a typical time of 1 fs, can recombine with a core hole rather than with the valence hole from that it was previously tunnel ionized and thus emit a much higher energy.

\begin{figure}[h]
\begin{center}
 \includegraphics[width=0.49\textwidth]{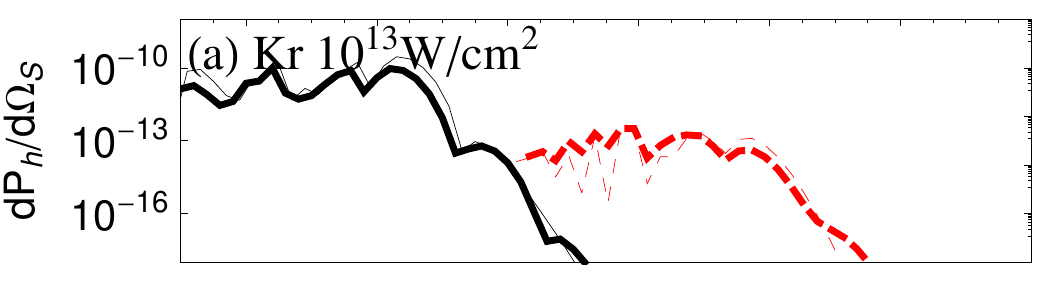}\hskip 0.2cm
 \includegraphics[width=0.49\textwidth]{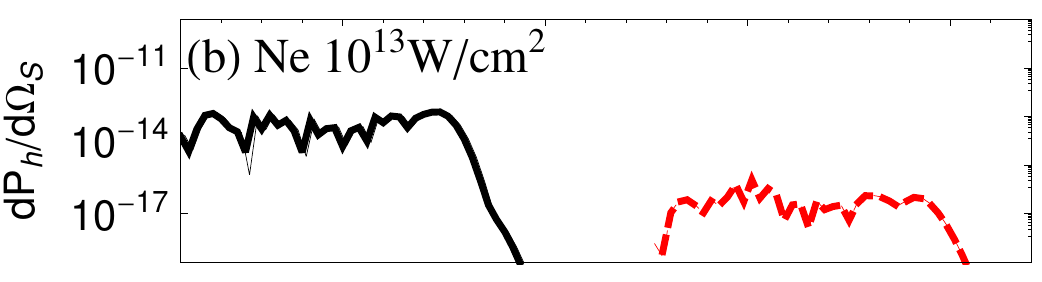}\\
 \includegraphics[width=0.49\textwidth]{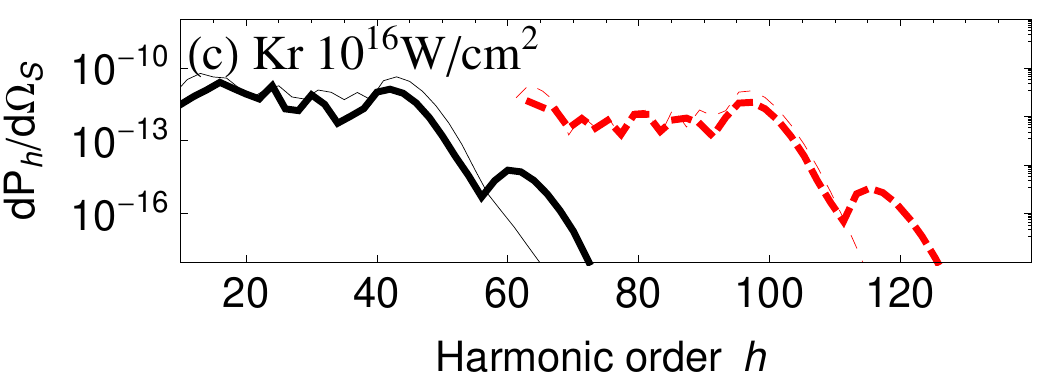}\hskip 0.2cm
 \includegraphics[width=0.49\textwidth]{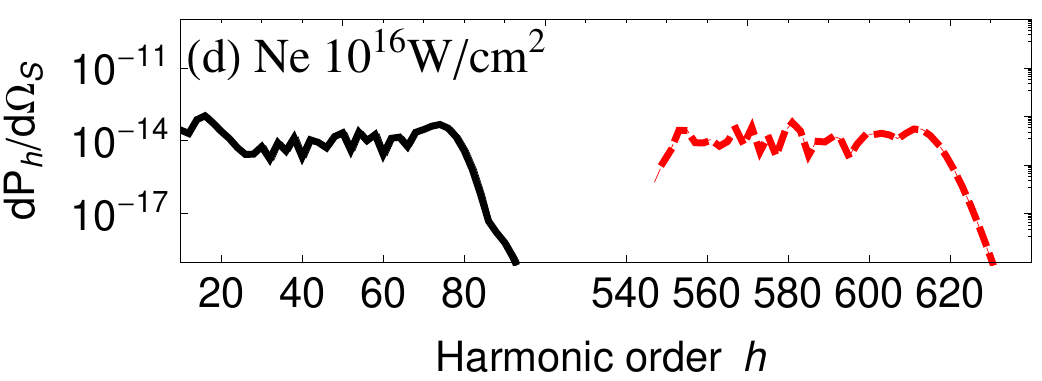}
\caption{HHG photon numbers of the $h^{\textrm{th}}$ harmonic for different  x-ray intensities from (a), (c) krypton and  (b), (d) neon. The solid black line stands for the recombination to the valence state whereas the more energetic plateau arising from core hole recombination is red the dashed line. The thin lines in the spectrum are obtained by neglecting ground-state depletion due to direct x-ray ionization. The xuv and laser pulse durations  are three optical laser cycles in all cases.} \label{fig:Rabi-cont}
\end{center}
\end{figure}

We developed an analytical formalism to cope with the two-electron two-color problem~\cite{buth:ra-11,buth:pra-11}. A two-electron Hamiltonian is constructed mostly from tensorial products of one-electron Hamiltonians that describe the 
%and similar assumptions as in~\cite{lewenstein} are employed. 
losses due to tunnel ionization and direct x-ray ionization via  phenomenological decay constants in conjunction with Auger decay of the intermediate hole. 
The system is described by equations of motions based on the Schr\"odinger equation and the solutions can be found in the dressed state basis. 
We apply our theory to the $3d \rightarrow 4p$ resonance in a krypton cation as well as to the $1s \rightarrow  2p$ resonance in a neon cation. The results for a resonant sinusoidal  x-ray field for two different intensities are shown in Fig.~\ref{fig:Rabi-cont}. The chosen optical laser field intensity is $3\times10^{14}\Wcm$ for krypton and  $5\times10^{14}\Wcm$ for neon both at $800\U{nm}$ wavelength.

The most striking feature in the obtained spectra is the appearance of a second plateau. It is upshifted in energy with respect to the first plateau by the energy difference between the two involved core and valence states. The two plateaus have comparable harmonic yields for x-ray intensities above $10^{16}\Wcm$. Quite importantly, the losses due to x-ray ionization do not lead to a significant drop of the HHG rate as can be seen by comparing the thick and thin lines. Both lines almost coincide for neon due to the negligible depletion induced by ionization. The second plateau bears signatures of the core state and may offer a route for ultrafast time-dependent chemical imaging of inner shells~\cite{ITATANI2004,morishita2008}. Moreover, by exploiting the upshift in energy, attosecond x-ray pulses come into reach.

\section{High-harmonic generation without attochirp}
\label{sec:attoHHG}
A particularly fascinating property of HHG is its time structure enabling for the generation of extremely short pulses~\cite{GOULIELMAKIS2008,SANSONE2010}. Nowadays pulses down to a duration of 63~as~\cite{KO2010} have been generated and the bandwidth to generate pulses of only 11~as is available~\cite{CHEN2010}. The emitted pulses have an intrinsic chirp, the so-called attochirp~\cite{MAIRESSE2003,KAZAMIAS2004} and, thus, are much longer than their bandwidth limit. To compensate the attochirp, dispersive optical media~\cite{KO2010,chirp-compensation} are employed. 

An alternative way to circumvent this problem would be to modify the HHG process such that the light is emitted without attochirp. 
%Therefore, we propose a scheme to engineer the attochirp by altering the harmonic generation process. 
It is shown~\cite{KOHLER:CF-11} that by means of laser pulse shaping, employing soft x rays for ionization~\cite{xuv-ass-HHG} and using an ionic gas medium, attosecond pulses with arbitrary chirp can be formed including the possibility of attochirp-free HHG and bandwidth-limited attosecond pulses. 

\begin{figure}%[h]
\begin{center}
\includegraphics[width=0.5\textwidth]{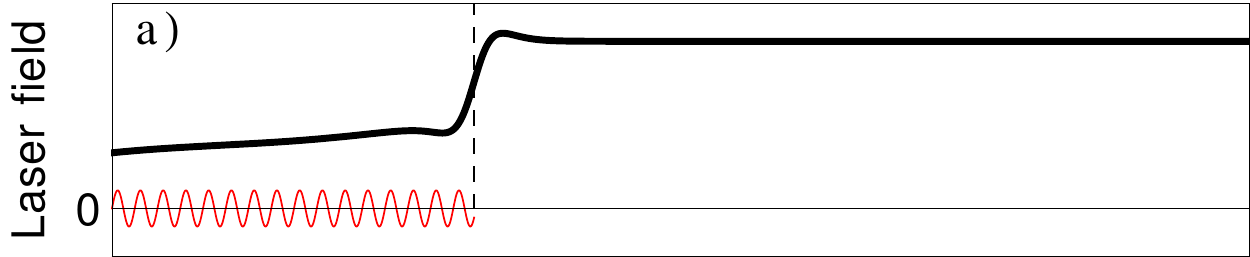}\\
\hskip0.0cm\includegraphics[width=0.5\textwidth]{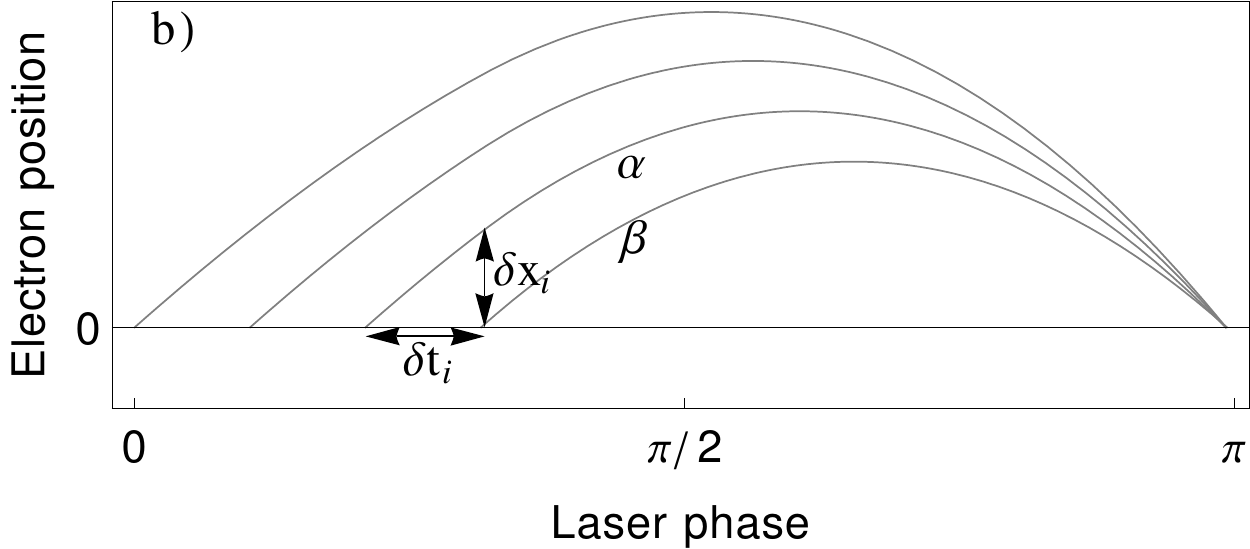}
\caption{Schematic of the recollision scenario: a) A half cycle of the tailored laser field (black). The red line is the assisting x-ray pulse. b) Different one-dimensional classical trajectories in the field of (a) which start into the continuum at different times but revisit the ionic core at the same time. Reproduced from Ref.~\cite{KOHLER:CF-11}.  Copyright (2011) by the Optical Society of America.} \label{figopti}
\end{center}
\end{figure}

The principles are illustrated in the trajectory picture of HHG~\cite{CORKUM1993,lewenstein} in Fig.~\ref{figopti}. Since the recollision time of a certain harmonic can be identified with its group delay~\cite{MAIRESSE2003,KAZAMIAS2004} in the emitted pulse, a simultaneous recollision of all trajectories would lead to a bandwidth-limited attosecond pulse. The demand of simultaneous recollision can be fulfilled if the electron is freed by single-photon ionization when the x-ray frequency $\omega\In{X}$ is much larger than the binding energy. In this case the electron has a large initial kinetic energy directly after ionization. Let us focus on the two example trajectories marked by $\alpha$ and $\beta$ in Fig.~\ref{figopti}b. Both are ionized at instants separated by a small time difference $\delta t\In{i}$.  The chosen starting direction of the trajectories along the laser polarization direction is such that they are subsequently decelerated by the laser field and eventually recollide. Note that the velocity difference between $\alpha$ and $\beta$ is conserved in time for a homogeneous laser field. With a convenient choice of the parameters, the velocity difference between both acquired during $\delta t\In{i}$ can be such that both recollide simultaneously.

\begin{figure}[th]
\begin{center}
\includegraphics[width=0.9\textwidth]{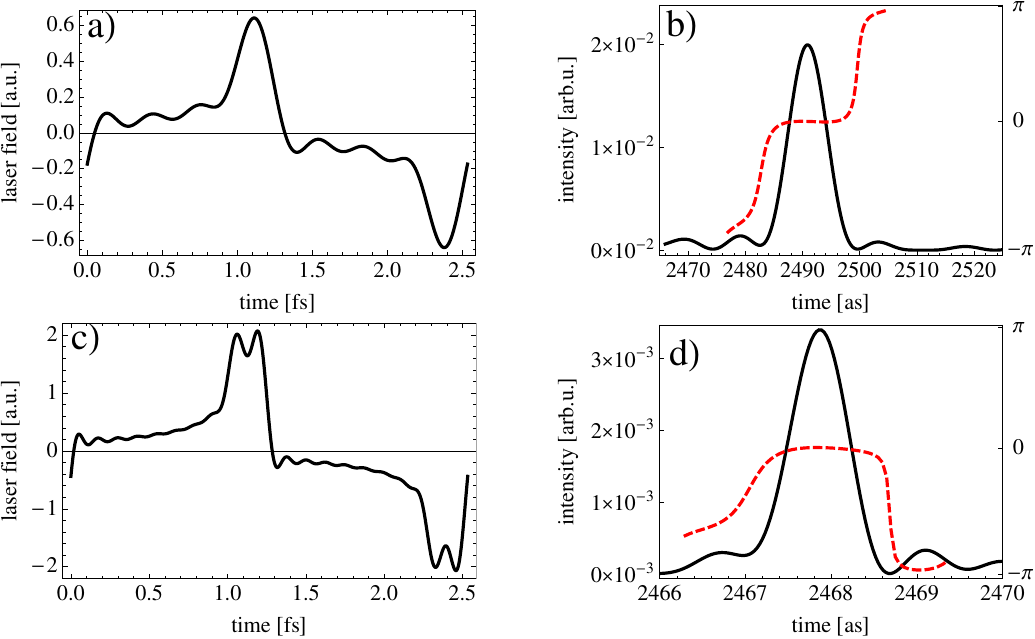}
\caption{a) the laser field composed of 8 Fourier components that illuminates an Li$^{2+}$ ion with $I\In{p}=4.5$~a.u.. The parameters of the additional x-ray field are indicated in the first row of Table \ref{tab_values}. The resulting 8~as pulse is shown in b). c) displays the laser field needed to create a pulse of 800~zs duration from Be$^{3+}$ ions with $I\In{p}=8$~a.u. The parameters are indicated in the second row of Table~\ref{tab_values}. The respective pulse is shown in d). The dashed red lines are the temporal phases of the pulses. Reproduced from Ref.~\cite{KOHLER:CF-11}.  Copyright (2011) by the Optical Society of America.} \label{fig_highE}
\end{center}
\end{figure}

In the following, we exemplify our method in two cases  producing bandwidth-limited pulses below 10~as and 1~as.
The optimized optical laser pulses (frequency $\omega=0.06$a.u.) are shown in Fig.~\ref{fig_highE}a and c and the parameters are indicated in Table~\ref{tab_values}.

The attosecond pulses are bandwidth limited as can be seen from the spectral phases (red lines in Fig.~\ref{fig_highE}b and d).

\begin{table}[h]
\begin{center}
\begin{tabular}{c|c|c|c|c|c}
$N_{F}$&$I_L$[W/cm$^2$]&$\omega_x$[eV]&$I_X$[W/cm$^2$]&ion&$I_p$ [a.u]\\
\hline
8&$10^{16}$&218&$3.5\times 10^{14}$&Li$^{2+}$&4.5\\
\hline
20&$10^{17}$&996&$1.4\times10^{15}$&Be$^{3+}$&8\\
\hline
\end{tabular}
\caption{Parameters for the two examples in Fig.~\ref{fig_highE}: $N\In{F}$ represents the number of Fourier components contained in the tailored driving pulse, $I\In{L}$ its peak intensity, $\omega\In{X}$  the x-ray frequency employed for ionization, $I\In{X}$ its intensity and the ionization energy $I\In{p}$.}\label{tab_values}
\end{center}
\end{table}

However, the method still has several drawbacks. It is experimentally demanding to create a pure ionic gas and to achieve phase-matching in a macroscopic medium due to the free electron dispersion.  Moreover, the required large initial momentum and the dipole angular distribution of the ionization process lead to an increased spread of the ionized wave packet as compared to tunnel ionization. Further, the ionization rate is small due to $\omega\In{X}\gg I\In{p}$. 
Precise shaping of intense driver fields  with intense harmonics as well as the synchronization of the x-ray and IR pulses is also demanding. A potential route to overcome the former difficulties could be to derive the x-rays, the IR light and its Fourier components from the same FEL electron bunch.

\section{Resonant photoionization involving two atoms}
\label{sec:ICD}
Interatomic electron--electron correlations are responsible for a variety of interesting phenomena, ranging from dipole--dipole interactions in cold quantum gases to F\"orster resonances between biomolecules. They may also lead to very characteristic effects in the photoionization of atoms. 

As an illustrative example, let us consider resonant two-photon ionization in a system consisting of two hydrogen atoms~\cite{MUELLER2011}. The atoms are assumed to be separated by a sufficiently large distance $R$ so that one may indeed speak about individual atoms. The system interacts with a monochromatic laser field whose frequency is resonant with the $1s$--$2p$ transition in hydrogen. In this situation there are two quantum pathways for ionization: (i) the direct channel where a single atom is ionized by absorbing two photons from the field, without any participation of the neighboring atom; and (ii) an interatomic channel where both atoms are first excited to the $2p$ level by absorbing one photon each and afterwards the doubly excited two-atom state decays via so-called interatomic Coulombic decay (see~\cite{ICD} for recent reviews). In other words, one of the atoms deexcites and transfers its energy radiationlessly to the second atom, this way causing its ionization. 

Leading to the same final state, the direct and two-center ionization pathways show quantum mechanical interference. It becomes manifest in the photoelectron spectra which consist of four lines due to the Autler-Townes splitting of the atomic levels in the external field~\cite{MUELLER2011}. Moreover, the two-center channel can be remarkably strong and even dominate over the direct channel by orders of magnitude. In fact, the ratio $[d/(R^3E)]^2$ determines the strength of the former with respect to the latter, where $d$ is the $1s$--$2p$ transition dipole moment and $E$ denotes the laser field strength. For example, at $R=1$\,nm and $E=10^4$\,V/cm this ratio is of the order of $10^4$.

Interatomic resonant photoionization may also occur in hetero-atomic systems~\cite{MARPE,NAJJARI2010}. In this case, one of the centers is first excited by single-photon absorption and subsequently, upon deexcitation, the partner atom is ionized. Evidently, this mechanism requires an hetero atomic system with an excitation energy of the one center exceeding the ionization potential of the other.

The interplay between resonant laser fields and interatomic electron-electron correlations can thus give rise to interesting and rather unexpected effects.
\\

C.B.~and M.C.K.~were supported by a Marie Curie International Reintegration
Grant within the 7$^{\mathrm{th}}$~European Community Framework Program
(call identifier: FP7-PEOPLE-2010-RG, proposal No.~266551).
C.B.'s work was funded by the Office of Basic Energy Sciences,
Office of Science, U.S.~Department of Energy, under Contract
No.~DE-AC02-06CH11357. T.P. acknowledges support by an MPRG grant of the Max-Planck Gesellschaft.
This work was supported in part by the Extreme Matter Institute EMMI.

%%%%%%%%%%%%%%%%%%%%%%%% referenc.tex %%%%%%%%%%%%%%%%%%%%%%%%%%%%%%
% sample references
% %
% Use this file as a template for your own input.
%
%%%%%%%%%%%%%%%%%%%%%%%% Springer-Verlag %%%%%%%%%%%%%%%%%%%%%%%%%%
%
% BibTeX users please use
% \bibliographystyle{}
% \bibliography{}
%

\end{document}